\documentclass{aastex61}

\received{~~}
\revised{~~}
\accepted{Feb 16, 2018}     
\submitjournal{ApJ}

\def\cep		{Cepheus~A}

\def\etal       {et~al.}
\def\muas     {~$\mu$as}
\def\kms       {~km~s$^{-1}$}

\begin{document}

\title{
Sun-Sized Water Vapor Masers in Cepheus~A}

\correspondingauthor{J.M.\ Moran}
\email{jmoran@cfa.harvard.edu}

\author{A.M.~Sobolev}
\affiliation{Ural Federal University, Ekaterinburg, Russia}

\author{J.M.~Moran}
\affiliation{Harvard-Smithsonian Center for Astrophysics, 60 Garden Street, Cambridge, MA 02138, USA}

\author{M.D.~Gray}
\affiliation{Jodrell Bank Centre for Astrophysics, School of Physics and Astronomy, Alan Turing Building, University of Manchester, M13~9PL, UK}

\author{A.~Alakoz}
\affiliation{Astro Space Center of the Lebedev Physical Institute, Moscow, Russia}

\author{H.~Imai}
\affiliation{Science and Engineering Area of the Research and Education Assembly, Kagoshima University, 1-21-35 Korimoto, Kagoshima 890-0065, Japan}

\author{W.A.~Baan}
\affiliation{ASTRON---Netherlands Foundation for Research in Astronomy, Dwingeloo, The Netherlands}

\author{A.M.~Tolmachev}
\affiliation{Astro Space Center of the Lebedev Physical Institute, Moscow, Russia}

\author{V.A.~Samodurov}
\affiliation{Astro Space Center of the Lebedev Physical Institute, Moscow, Russia}
\affiliation{National Research University, Higher School of Economics, Moscow, Russia}

\author{D.A.~Ladeyshchikov}
\affiliation{Ural Federal University, Ekaterinburg, Russia}


\begin{abstract}

We present the first VLBI observations of a Galactic water maser (in \cep) made with a very long baseline interferometric array involving the RadioAstron Earth-orbiting satellite station as one of its elements. We detected two distinct components at --16.9 and 0.6\kms\ with a fringe spacing of 66~microarcseconds ($\mu$as). In total power, the 0.6\kms\ component appears to be a single Gaussian component of strength 580~Jy and width of 0.7\kms. Single-telescope monitoring showed that its lifetime was only 8~months. The absence of a Zeeman pattern implies the longitudinal magnetic field component is weaker than 120~mG. The space--Earth cross power spectrum shows two unresolved components smaller than 15\muas, corresponding to a linear scale of $1.6\times10^{11}$~cm, about the diameter of the Sun, for a distance of 700~pc, separated by 0.54\kms\ in velocity and by $160\pm35$\muas\ in angle. This is the smallest angular structure ever observed in a Galactic maser. The brightness temperatures are greater than $2\times10^{14}$~K, and the line widths are 0.5\kms. Most of the flux (about 87\%) is contained in a halo of angular size of $400\pm150$\muas. This structure is associated with the compact HII region HW3diii. We have probably picked up the most prominent peaks in the angular size range of our interferometer. We discuss three dynamical models: (1) Keplerian motion around a central object, (2) two chance overlapping clouds, and (3) vortices caused by flow around an obstacle (i.e., von~K\'arm\'an vortex street) with Strouhal number of about~0.3.

\end{abstract}




\section{Introduction} \label{sec:intro}

\begin{figure}[hb!]
\epsscale{0.6}
\plotone{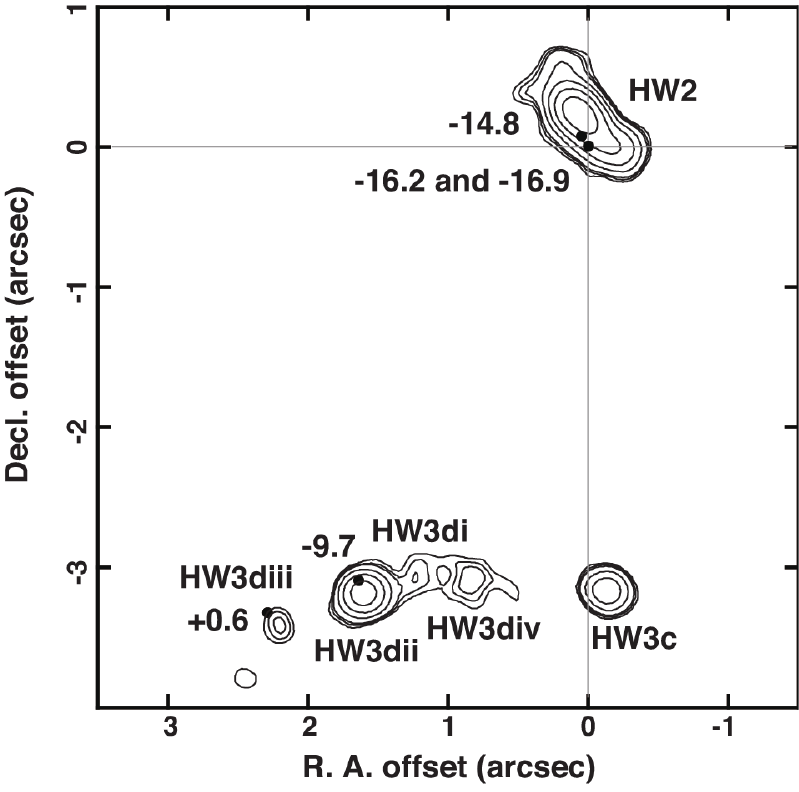}
\caption{The central part of the star-forming region \cep. The contours show the extent of the continuum components taken from the 1.3~cm VLA image [adapted from \citet{Torrelles98}]. The nomenclature is based on the original identification of about 16 continuum radio sources marking the sites of newly formed massive stars by \citet{HughesWouterloot84}. The dots mark the positions of masers (labeled by their velocities) whose positions were found by analysis of the relative fringe rates derived from these observations. The coordinate origin is the center of HW2/R4: ${\rm RA}=22^{\rm h}56^{\rm m}17.977^{\rm s}$, ${\rm dec}=61^{\rm d}45'49.37''$ (2000). The relative alignment of the masers and continuum is accurate to about $\pm0.2''$ (see text). At a distance of 700~pc, $1''$ corresponds to $1.05\times10^{16}$~cm.}
\label{fig:f1}
\end{figure}

\cep\ is a region of massive star formation within our Galaxy. Its radio continuum image consists of about 16 compact thermal cores, many of which are associated with embedded heating sources in the form of newly formed O and B stars. These sources were first identified by \citet{HughesWouterloot84} and are numbered with the prefix HW. The distance to the complex has been determined to be $700\pm40$~pc both by VLBI parallax measurement from the continuum emission from HW9 \citep{Dzib11} and methanol masers associated with HW2 \citep{Moscadelli09}. HW2 is the dominant energy source in the complex. Its continuum emission arises from an elongated structure (see Fig.~\ref{fig:f1}), which has been identified as a thermal jet with an outflow velocity of 480\kms\ \citep{Curiel06}. Another structure, perpendicular to the jet, is a disk of dust and molecular gas \citep{Patel05}. A system of water masers is associated with this disk, whose components are spread over an area of about $0.5''$ in extent \citep{Torrelles98}. Another important source is HW3, which lies about $3''$ south of HW2. The radio continuum emission shows four distinct cores, all probably associated with newly formed B stars \citep{Hughes95}. Most of the water masers associated with HW3d define a highly collimated outflow centered on HW3dii \citep{Chibueze12}. Of particular interest to this study is the source HW3diii, which lies about $0.5''$ east of HW3dii. The morphology of the HW2 and HW3 regions is shown in Fig.~\ref{fig:f1}. For a general discussion of the physics of cosmic masers, see \citet{Gray12}.

\begin{figure}[ht!]
\epsscale{0.5}
\plotone{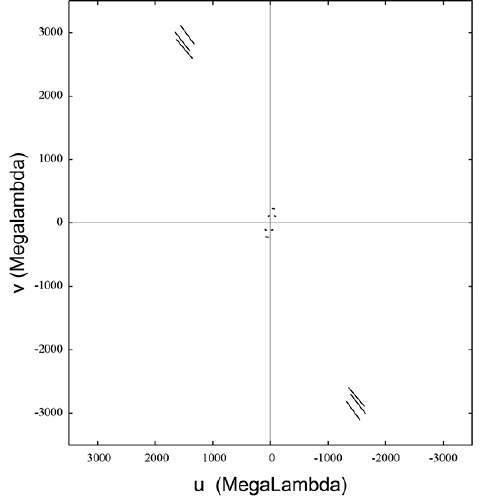}
\caption{$(u,v)$ plane coverage of the 40-minute observation of \cep\ on 18~Nov 2012 in millions of wavelengths.}
\label{fig:f2}
\end{figure}

We present in this paper our measurements of the maser emission from Cepheus A made with an unprecedented resolution (at the time of observations)  of 66~microarcseconds ($\mu$as)  on a baseline of 3.3 Earth diameters (ED). These are among the earliest results from a VLBI experiment that incorporate the RadioAstron satellite radio telescope (SRT). More recently, observations with baselines up to 10~ED on other galactic masers and up to 26.7~ED on extragalactic masers have been presented in conference proceedings \citep{Sobolev18, Shakhvorostova18, Baan18}. The only other reported detection of an H$_2$O maser with a space VLBI experiment was of the very bright  maser the  in Orion-KL region, but with projected baseline shorter than an Earth diameter \citep{Kobayashi00}.

The properties of the SRT, which was launched in 2011, are described by \citet{Kardashev13} and \citet{RadioAstron18}. The SRT operates at frequencies of 22, 5, 1.6, and 0.3~GHz. The receiving element is a 10-m parabolic dish, whose aperture efficiency is about 10\% at 22~GHz. The local oscillator phase is controlled by an onboard hydrogen maser. There are four  baseband channels:  two subbands of 16 MHz in  each sense of circular polarization. These signal streams were digitally sampled with one-bit quantization, transmitted to Earth and recorded for later processing at the VLBI center in Moscow. 

\section{Observations} \label{sec:observ}

The observations were made in a single 40-minute period from 12:00~UT to 12:40~UT on 18~Nov 2012. The data were blocked into four segments of 600-second duration each. The actual observation time on each segment was 570~seconds. The VLBI array consisted of the SRT and ground-based telescopes at Yebes, Spain (Ys); Noto, Italy (Nt); and Zelenchukskaya, Russian Federation (Zc). The diameters of these telescopes are 40, 32, and 32~m, respectively. Over the 40-minute observation, the $(u,v)$ coordinates of the SRT--Ys baseline changed from (1.36,~2.60) to (1.63,~2.89) in units of Giga-wavelengths. The corresponding fringe spacings changed from 70 to 62\muas\ (corresponding to 0.049 and 0.043~AU, or 7.3 and $6.4\times10^{11}$~cm, respectively). The mean position angle of the space--Earth baseline was $28^\circ$. The $(u,v)$ coverage for the full 40 minutes is shown in Fig~\ref{fig:f2}. The data were correlated using the Astro Space Center (ASC) software correlator \citep{Likhachev17},   but only two spectral subsets of the data, which contained all {known spectral components were retained  (one 8-MHz subband in each polarization). The post-correlation data reduction, including fringe fitting, was carried out with the PIMA calibration package \citep{Petrov11}. Most subsequent analysis was carried out with new ad~hoc software suitable for space VLBI data.} The processing configuration provided 1024 channels, resulting in a channel spacing of 7.81~kHz, corresponding to 0.105\kms. The final processing was completed after the determination of the best orbital parameters for the SRT, which were accurate to 500~m in position and 0.02~m~s$^{-1}$ in velocity \citep{Stepanyants17}.

\begin{figure}[ht!]
\epsscale{0.7}
\plotone{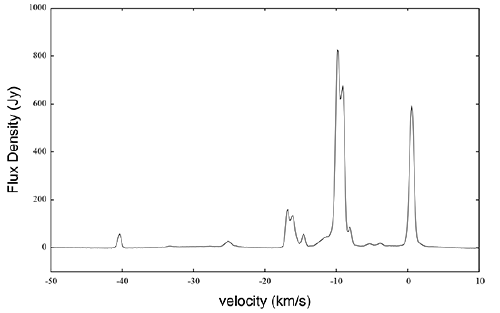}
\caption{The total power spectrum (average of the RCP and LCP spectra) from the first 600-second segment of observations at the Yebes telescope. No off-source reference spectrum was available, so a polynomial baseline was fit to the signal-free parts of the spectrum and removed. The velocity is with respect to the local standard of rest (LSR). $V({\rm LSR})=0$\kms\ corresponds to $V({\rm heliocentric})=-7.5$\kms. On the 3.3 ED baselines between the SRT and ground stations, fringes were detected only on the --16.9 and 0.6\kms\ features.}   
\label{fig:f3}
\end{figure}

\begin{deluxetable}{ccccl}[ht!]
\tablecaption{Positions of H$_2$O masers in Cepheus~A\tablenotemark{a} \label{tab:t1}}
\tablecolumns{4}
\tablenum{1}
\tablewidth{0pt}   
\tablehead{
\colhead{}& \colhead{} & \colhead{} & \colhead{}&\colhead{~Continuum} \\
\colhead{Velocity (lsr)} & \colhead{$\Delta$ RA} & \colhead{$\Delta$ dec} & \colhead{Flux density} & \colhead{association} \\
\colhead{(km~s$^{-1}$)} & \colhead{($''$)} & \colhead{($''$)} & \colhead{(Jy)} &\colhead{~} \\
}
\startdata
\phn\phn0.6 & 2.29 & --3.33 & 580 & \phn\phn HW3diii \\
\phn--9.7 & 1.64 & --3.10 & 800 & \phn\phn HW3dii \\
--14.8 & 0.05 & \phn0.07 & \phn55 & \phn\phn HW2 \\
--16.2 & 0.01 & \phn0.01 & 130 & \phn\phn HW2 \\
--16.9 & 0\phd\phn\phn & \phn0\phd\phn\phn & 152 & \phn\phn HW2 \\
\enddata
\tablenotetext{a}{Relative position accuracy is $\pm0.2''$.}
\end{deluxetable}

\section{Results} \label{sec:results}

The total power spectrum obtained from the Yebes data is shown in Fig.~\ref{fig:f3}. Strong fringes were detected on all three ground baselines but only on the space baseline SRT--Ys. Weaker detections were achieved on the other space baselines but were not used in this analysis. The sensitivity of the cross power spectra was limited by the coherence time of the interferometer, which was about 100~seconds. We measured the fringe rates on the three ground baselines of the spectral features at 0.6, --9.7, --14.8, and --16.2, all with respect to the feature at --16.9\kms. We used the  task FRMAP in the Astronomical Image Processing System (AIPS)  described by \citet{Walker81} and \citet{Thompson17} to find the relative feature positions from their relative fringe rates. Each relative fringe rate localized the relative position of the feature to a line in RA--dec space. Although the hour angle spread provided by the 40-minute observation was small, the three  ground-based baselines provide a good spread in position angle such that accurate relative coordinates were obtained with an uncertainty of $\pm0.02''$ in each coordinate. The positions are listed in Table~\ref{tab:t1}. However, it is difficult to align the masers with the continuum. We have placed the --16.9\kms\ feature near the center of the outflow in HW2. The absolute positions of 39 masers associated with HW2 in 1995 were reported by \citet{Torrelles98}. None of these velocity components can be reliably associated with our detections. However, most of the strong components identified in 1995 were within $\pm0.3''$ of the center of~HW2. In particular, components near our velocity commonly appear in maser complex R4 \citep{Torrelles11}. We adopt $\pm0.2''$ as our alignment accuracy.

\begin{figure}[ht!]
\epsscale{0.5}
\plotone{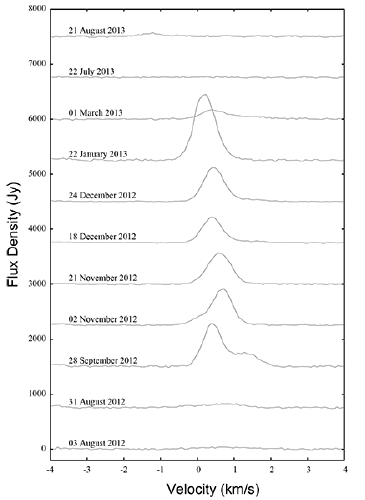}
\caption{Spectrum of the feature near 0.6\kms\ observed at the Pushchino Observatory from 3~Aug 2012 to 21~Aug 2013. Epochs of observation are shown by the dotted horizontal lines. Note the drift in the central velocity.}
\label{fig:f4}
\end{figure}

\begin{figure}[ht!]
\epsscale{0.85}
\plotone{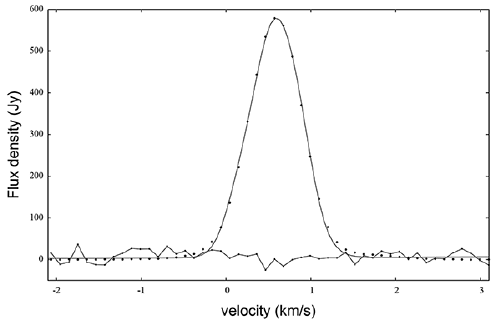}
\caption{Closeup of the 0.6\kms\ feature. The dots are the total power spectrum obtained from the Yebes telescope data (see full spectrum in Fig.~\ref{fig:f3}). The smooth line is a Gaussian profile fitted to the data. The straight line segmented curve is the difference between the RCP and LCP total power spectra after removing a gain factor. The scale of the difference spectrum has been multiplied by a factor of 20. The absence of any significant signal indicated that the magnetic field is less than 120~mG.}
\label{fig:f5}
\end{figure}

\begin{figure}[ht!]
\epsscale{0.8}
\plotone{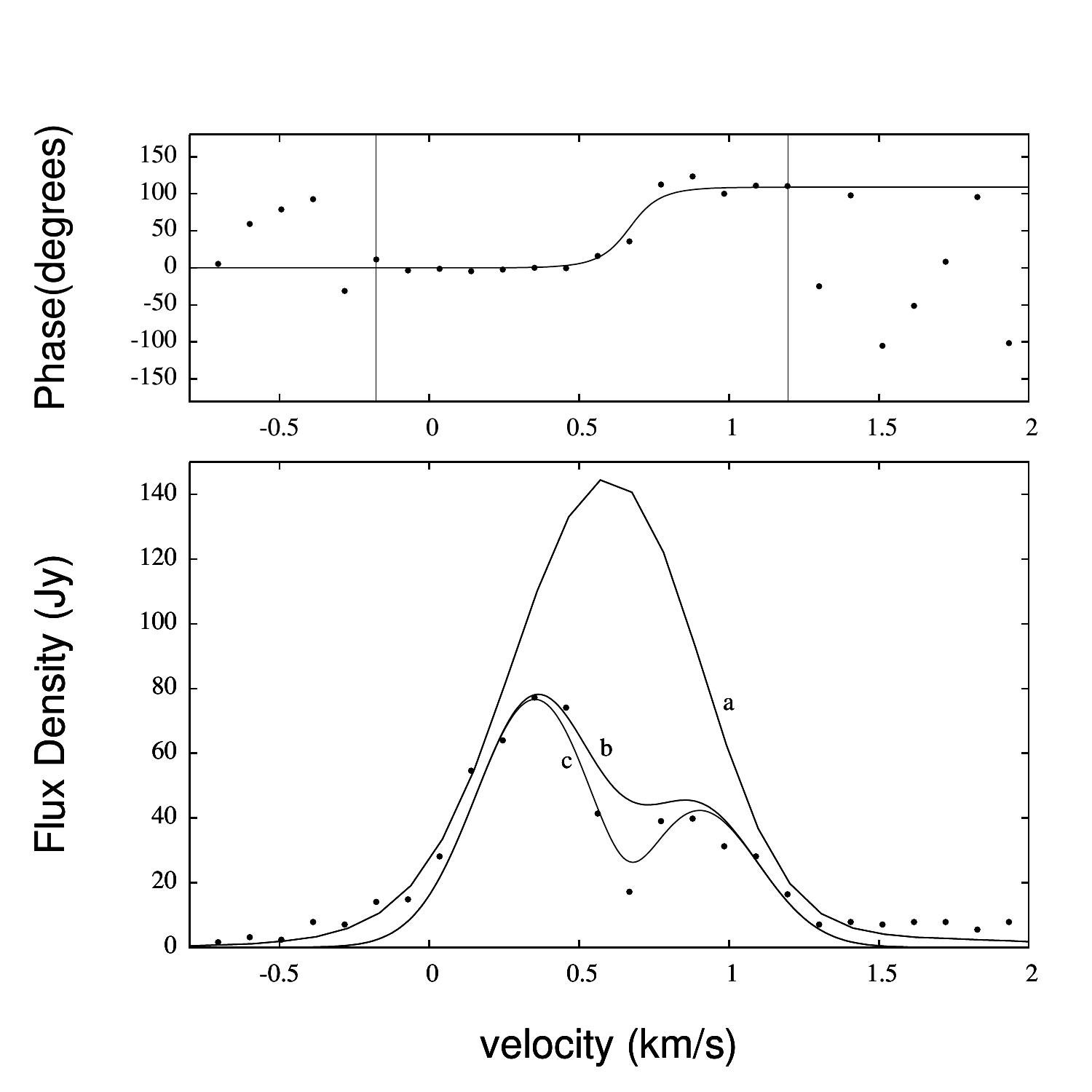}
\caption{The cross power spectrum (dots) from the SRT--Ys baseline data for the first 600-second block in LCP. The visibility phase and amplitude data are shown in the top and bottom plots, respectively. A complex two-component Gaussian model was fit to this data. This model is shown by the solid line in the top plot (phase), with the velocity range marked by the vertical lines within which the signal-to-noise ratio is adequate, and by curve (c) in the bottom plot (amplitude). Curve (b) is the scalar sum of the two spectral components, and curve (a) is the total flux density reduced in scale by a factor of  four for comparison.}
\label{fig:f6}
\end{figure}

Fringes on the SRT to ground baselines were only detected on the features at --16.9 and 0.6\kms\ (the detection threshold is about 2~Jy). The one at --16.9 \kms, which is associated with HW2, had a fringe visibility amplitude of only about 0.02. We focused our analysis on the strong isolated feature at 0.6\kms. Routine monitoring of the spectrum at the Pushchino Observatory indicates that most features persist for about a year. In particular, the feature at 0.6\kms\ appeared between 30~Aug 2012 and 20~Sept 2012 and disappeared between 1~March 2013 and 18~July 2013  (see Fig.~\ref{fig:f4}). Except for this time range, no features near this velocity were detected during the monitoring observations from 10~Oct 2010 to 14~Aug 2014. The rms noise level was typically 5~Jy. Thus, we can assign it a lifetime of $8\pm3$~months. The total power spectrum from our observations at Yebes is shown in Fig.~\ref{fig:f5}. A single Gaussian profile fit to the Stokes I spectral data (RCP~+~LCP)/2 gives the parameters: amplitude = $580\pm3$~Jy, velocity = $0.58\pm 0.01$\kms, and width (full width at half-maximum, FWHM) = $ 0.672\pm0.005$\kms. To search for the circular polarization, we calculated the Stokes V profile via the formula $V=S({\rm RCP})-a\times S({\rm LCP})$. The parameter $a$ accounts for the small unknown gain difference between the two polarizations and was chosen to minimize the mean square deviation between $S$(RCP) and $S$(LCP). A longitudinal component of  magnetic field in the maser medium will shift the profiles slightly in frequency. In this case, the $V$ profile has a distinctive shape proportional to the derivative of the total intensity profile. This is an anti-symmetric ``S"-shaped curve. The magnitude of the curve, $V_{\rm max}$, is related to the longitudinal component of the magnetic field by the equation \citep{FiebigGusten89}
\begin{equation}
V_{\rm max}/I_{\rm max}=13.4\times10^{-6}B/\Delta v~~,
\end{equation}

\begin{deluxetable}{cccC}[ht!]
\tablecaption{Visibility components of the 0.6\kms\ feature \label{tab:t2}}
\tablecolumns{4}
\tablenum{2}
\tablewidth{0pt}
\tablehead{
\colhead{Velocity} & \colhead{$S$} & \colhead{$\Delta v$} & \colhead{${T_B}^*$} \\
\colhead{(km~s$^{-1}$)} & \colhead{(Jy)} & \colhead{(km~s$^{-1}$)} & \colhead{(K)} 
}
\startdata
0.895 & 43 & 0.47 & 1.5\times10^{14} \\
0.355 & 77 & 0.47 & 3\phd\phn\times10^{14} \\
\enddata
\tablenotetext{*}{Lower limit.}
\end{deluxetable}

\begin{figure}[ht!]
\epsscale{0.8}
\plotone{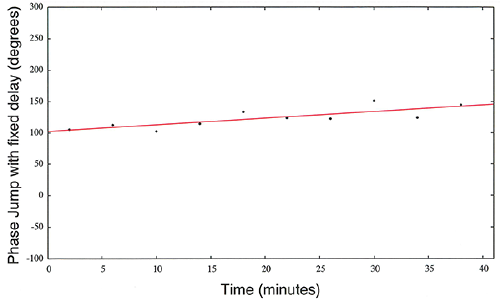}
\caption{The relative phase between the 0.895 and 0.355\kms\ subcomponents as a function of time during the 40-minute observation. The data have been coherently averaged to 4~minutes. The relative phase and relative phase drift over the observation of $45^\circ$ can be used to constrain the separation of the components. The orbit specification for RadioAstron of 0.02~m~s$^{-1}$ would allow a maximum of $\pm2^\circ$ of the observed phase shift to be caused by the change in the baseline error. AGN observations near the time of these observations suggest that the actual error is about four times smaller.}
\label{fig:f7}
\end{figure}

\noindent where $B$ is the line-of-sight magnetic field strength in mGauss (mG) and $\Delta v$ is the line width in\kms. We assumed the Zeeman parameters for the strongest hyperfine component of the 22~GHz transition  with $\Delta v= 0.672$\kms. There is no hint of a Zeeman signature at the level of 1.4~Jy ($V_{\rm max}/I_{\rm max}<2.4\times10^{-3}$). Hence, the line-of-sight component of the magnetic field strength is less than about 120~mG. For comparison, \citet{Vlemmings06} measured the magnetic fields in about 30 features in \cep, mostly in the HW2 region, and found them typically be in the range of 100--600~mG. 

The vector-averaged cross power spectrum of the 0.6\kms\ feature is shown in Fig.~\ref{fig:f6}. The spectrum shows two components with a sharp change in phase between them. This is a clear indication of a double source structure. We fit a double Gaussian profile to the complex cross power spectrum. The parameters of this fit are listed in Table~\ref{tab:t2}. We were not able to obtain a stable three Gaussian component fit to the total power spectrum. However, we believe the total power associated with the two components cannot be significantly greater than the cross power amplitudes or they would be clearly visible in the total power spectrum (see fitted profile in Fig.~\ref{fig:f5}). Hence, we assign both of them  visibility amplitudes of greater than 0.8, and hence  sizes of  less than 15 \muas\ which leads to the estimate of the lower limits of brightness temperature  in Table~\ref{tab:t2}.   Note that the normalized fringe visibility can be accurately determined because the total power spectrum can be measured with both the SRT and Ys telescope. In this case, the fringe visibility is simply the cross power spectrum divided by the geometric mean of the total power spectra in raw correlator units. Individual values of system equivalent flux densities (SEFD) from a~priori measurements are not needed. The fraction of flux (Jy\kms) in the cross power spectrum is $0.13\pm0.02$ of the 0.6\kms\ complex. This fraction is the ratio of the integrals of curve~b and curve~a in Fig.~\ref{fig:f6}. To further investigate the structure of the 0.6\kms\ component, we examined the cross power spectra on the three ground-only baselines. A careful calibration of the cross power spectra with the associated autocorrelation spectra on a minute-by-minute basis shows that the normalized fringe visibilities are 0.83, 0.61, and 0.53 for Ys--Nt, Nt--Zc, and Ys--Zc baselines of length 115, 138, and 228~M$\lambda$, respectively (see Fig.~\ref{fig:f2}). As mentioned above, the statistical uncertainty in these estimates is small because the system temperatures and telescope collecting areas drop out of the calculation, but the visibilities could be underestimated because of local oscillator coherence loss factors. These visibilities can be modeled approximately by a circular Gaussian disk of diameter (FWHM) of $400\pm15$\muas\ and flux density of 580~Jy. We refer to this structure as a halo. Note that we could not determine the registration between the halo and the compact double structure. The visibilities vs.\ baseline length and a cartoon of the maser components are shown in Fig.~\ref{fig:f9}.

\begin{figure}[ht!]
\epsscale{0.8}
\plotone{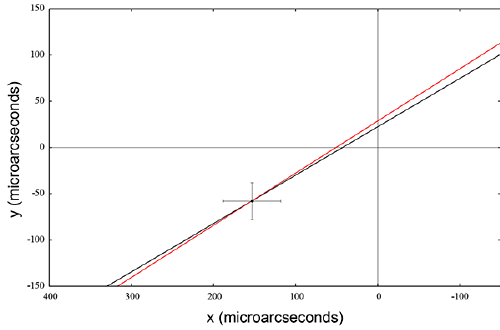}
\caption{The offset between the 0.90 and 0.36\kms\ subcomponents determined from relative phase measurements on the SRT--Ys baseline at 12:00  (red line) and 12:40 UT  (black line). Each measurement constrains the relative position to a line in position space.}
\label{fig:f8}
\end{figure}

\begin{figure}[ht!]
\epsscale{1.2}
\plotone{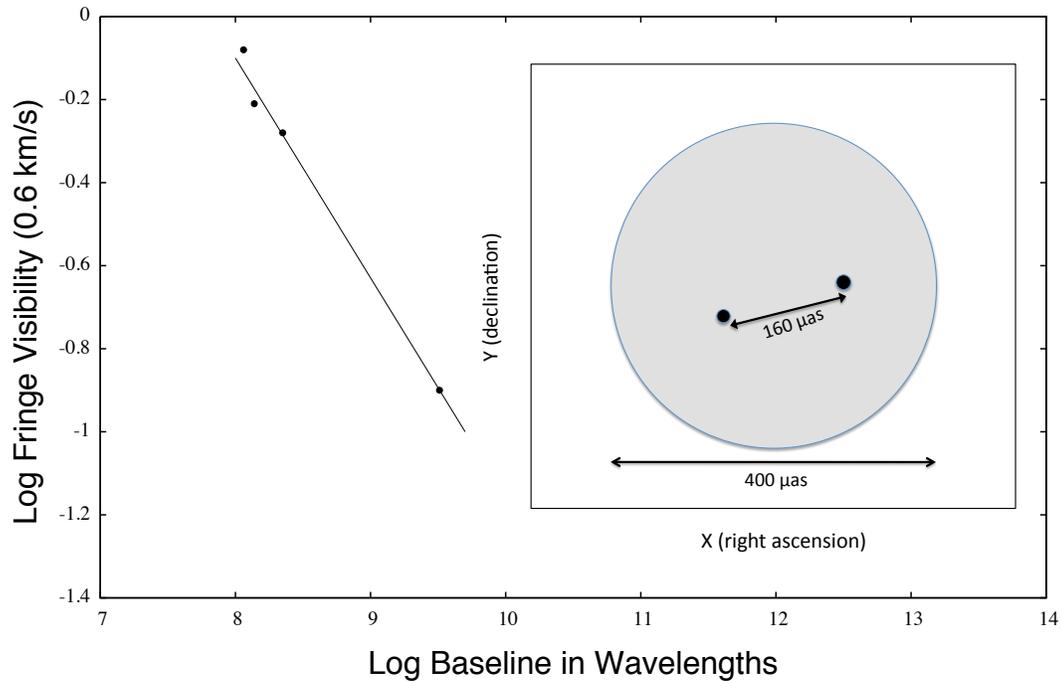}
\caption{The fringe visibility amplitude for the 0.6\kms\ feature on the three ground baselines vs.\ baseline length. The solid line is a model of a circular Gaussian halo of 400\muas\ angular diameter (FWHM) containing 96\% of the integrated flux density, plus an unresolved component to account for unresolved flux at large projected baselines. Inset: A cartoon of the maser emission from the 0.6\kms\ feature. The small components are modeled on the SRT--Ys baseline data, which show two subcomponents separated by 160\muas\ at a PA of $113^\circ$. This PA corresponds to the axis of the flow from Hd3ii. About 13\% of the integrated flux density is in the subcomponents. The subcomponents are shown centered on the halo, but this relative alignment is unknown.}
\label{fig:f9}
\end{figure}

The phase difference between the two  components of the 0.6\kms\ feature is about $125^\circ$ at the midpoint of the observations or about 0.35 of the fringe spacing, or 24\muas. If the features were aligned along the direction of maximum resolution at a PA of $28^\circ$ (see Fig.~\ref{fig:f2}), then they would be spaced by 24\muas. This is the minimum possible spacing. The actual separation and position angle can be estimated by the change in the relative phase of the features over the observations, which is $43^\circ$ (see Fig.~\ref{fig:f7}). The maximum contribution to this relative phase due to a change in instrumental delay caused by a baseline error is $\pm2^\circ$ \citep{Stepanyants17}. We thus are able to calculate a phase difference for the beginning of the observation to be $102\pm10^\circ$ and the phase difference at the end of the observation to be $145\pm10^\circ$. The position offset and its PA can be determined by the two $(u,v)$ plan measurements, as shown in Fig.~\ref{fig:f8}. The baseline rotates by only about $3^\circ$, but this is sufficient to determine the offset to be $160\pm35$\muas\ at a PA of $113\pm5^\circ$. This corresponds to a projected velocity gradient of 4\kms~AU$^{-1}$. 

The question arises as to whether the size estimates of the components could be affected by interstellar scattering. The angular broadening of images due to the turbulent interstellar medium can be estimated from the NE2001 model of \citet{CordesLazio03}. For the Galactic longitude of $109.8^\circ$ and latitude of $2.1^\circ$, the integrated effect over 700~pc at 22.2~GHz is 7\muas. Hence, scattering could only have a small effect on our measurements.

\section{Discussion} \label{sec:discuss}

Three properties of the 0.6\kms\ feature clearly distinguish it from the other \cep\ features detected in our observations: (1) the doublet structure of the feature revealing itself only at the space--ground baselines, (2) the unusual value of the radial velocity (i.e., a value not prominently represented in the cluster of masers near HW2 and HW3dii), and (3) its strong variability with nonlinear drift in velocity with time (see Fig.~\ref{fig:f4}). As discussed in the first subsection below, the existence of the doublet structure can have a spectroscopic explanation, but evidence also exists that the real explanation is astrophysical in nature, as discussed in the further subsections. The most likely explanation of the structure is that it results from turbulence on a variety of scales up to 400\muas. The two peaks we detected may be simply the emission peaks on the principal scale that the SRT--ground baselines are sensitive to, i.e., tens to hundreds of\muas.

In order to understand the physical nature of the 0.6\kms\ feature, we need to determine the type of astrophysical object associated with it. Results presented in Table~1 and Fig.~\ref{fig:f1} show that emission in this feature comes from the area around the compact HII region HW3diii (see Fig.~\ref{fig:f1}). Maps by \citet{Chibueze12} show that the widespread maser features corresponding to the outflow in this area have proper motion velocities around 10\kms. The features with the other velocities, including more redshifted ones, are located in the turbulent central maser cluster. The presence of a circumstellar disk or envelope around a young stellar object (YSO) can explain the relatively high range of velocities observed. Results obtained by \citet{Chibueze12} provide strong support to the presence of the massive YSO in the region. The very close location of the most redshifted maser feature to that of the most blueshifted one makes the disk hypothesis more likely. 

Turbulent motions in the form of evolving 3D vortices (eddies) are characteristic of the environment of massive YSOs. The turbulence is introduced at the largest scales determined by the boundary conditions and dissipates at the smallest scales determined by viscosity. The turbulence is generated in the circumstellar disks around massive YSOs, where it plays a decisive role in the mixing of material, momentum transfer, and other processes important for the disk structure and evolution. Excellent theoretical examples of the turbulent vortex formation in accretion disks can be found in, e.g., \citet{Meheut10} and \citet{KurBisKay14}. Unfortunately, manifestations of a turbulent vortex in maser emission from accretion disks are much less studied observationally. The main problem is the difficulty of associating maser sources with their locations in the disks. This has been addressed only in a few cases [e.g., \citet{Gallimore03} for the R4 maser arc near \cep\ HW2, \citet{Sanna17} for the \cep\ HW2 disk, and \citet{Sanna15} for G023.01-00.41].

In contrast, in the outflows from massive YSOs, the largest and intermediate scales of turbulence are well traced by water maser observations [see the papers on W49N by \citet{Walker84} and \citet{Gwinn94a} and more recent papers on the nearby sources \cep\ and W75N \citep{Uscanga10} and W3IRS5 \citep{Imai02}. In their consideration of the maser data on the turbulence in the flows from the massive YSOs, \citet{Strelnitski02} proposed that ``the maser hot spots originate at the sites of ultimate dissipation of highly supersonic turbulence." This assertion finds support in a well-ordered spatio-kinematical pattern in the small-scale water maser features reported in \cep\ HW2 by \citet{Uscanga03}, in W75N by \citet{Uscanga05}, and in W49N by \citet{Gwinn94b}.  Observations of \citet{Uscanga05} suggested microstructure with a size about 1~AU. This structure had a short lifetime supposedly on the order of a month. All information on the known examples of the structures with an ``eddy-like" spatio-kinematical pattern does not contain evolutionary information and has a form of snapshots, although the other maser structures in \cep\ HW2 show persistence on the time scales of years \citep{Torrelles01}. \citet{Uscanga05} speculated that these short-lived kinds of spatio-kinematical microstructures are ‘’either produced by fluid instabilities within the shocked material or correspond to nearly round cloudlets (turbulent eddies?) in the ambient medium.

In the sections below, we discuss a spectroscopic origin for our observations as well as three dynamical phenomena that may explain them.

\subsection{ Spectroscopic Origin: Hyperfine Splitting} \label{sec:hyperfine} 

The velocity separation of 0.54\kms\ between the components in Fig.~\ref{fig:f6} is close to the velocity separations of the H$_2$O hyperfine splittings of 0.45\kms\ between the $F=7$--6 and $F=6$--5 transitions and 0.58\kms\ between $F=6$--5 and $F=5$--4. However, there are several problems with this spectroscopic hypothesis. First, the components of the double-peaked spectrum in Fig.~\ref{fig:f6} are not spatially coincident, so they would have to be associated with different hyperfine components. Second, the $F=7$--6 hyperfine transition has the lowest frequency (and $F=5$--4 the highest) of the three strong transitions \citep{Kukolich69}, while the strength order is $F=7$--6, $F=6$--5, then $F=5$--4 \citep{DeguchiWatson86}. We would therefore expect the strongest peak at the lowest frequency (most positive Dopper shift) in our spectrum, but the opposite is seen in Fig.~\ref{fig:f6}. Moreover, all three of the hyperfine transitions introduced above have comparable line strengths [see Fig.~1 of \citet{DeguchiWatson86}], so a triplet spectrum would be expected rather than a doublet in the case of hyperfine intensity anomalies. We think this explanation is unlikely because it would require some complicated combination of hyperfine-specific pumping and/or competitive gain effects to generate the observed spectrum. 

\subsection{Keplerian Rotation} \label{sec:keplerian} 

For the first  dynamical interpretation, the maser hot spots might (see Fig.~\ref{fig:f9}) be amplifying along chords (i.e., filaments) in the plane of a Keplerian disc, orbiting a protostellar or protoplanetary object, viewed approximately edge-on. In this case, the length of the filaments responsible for the emission is displaced radially by 80\muas\ ($8.4\times10^9$~m) from the center. A rotational velocity equal to 0.27\kms, half the velocity separation of the components, gives a central mass, $M=rv^2/G$, of $9.1\times10^{24}$~kg, or approximately 1.5 Earth masses. The orbital period would be 2300 days, which is much longer than our monitoring period. A very large maser depth (negative optical depth) is possible in a disc if the number density is close to the maximum for strong collisional pumping of the 22~GHz transition: $n=2\times10^{10}$~cm$^{-3}$ at $T_K=750$~K in largely dust-free gas \citep{Gray16}. Under these conditions, a 1\% inversion with an ortho-H$_2$O abundance of $3\times10^{-5}$ yields a gain coefficient of $1.05\times10^{-8}$~m$^{-1}$, and therefore a maser depth would be well above the level needed to achieve saturation. Under this hypothesis, the splitting of the 0.6\kms\ feature can be explained by rotation of the planetary object around the massive YSO in the region.

\subsection{A Pair of Approximately Spherical Clouds} \label{sec:2clouds} 

The second dynamical interpretation is that the 0.6\kms\ maser emission results from the partial overlap, along the line of sight, of a pair of approximately spherical clouds. This alignment could be random, although it is much more likely that the objects are related.  The clouds may have a very large relative velocity, provided that the dominant component lies in the plane of the sky. The relative velocity along the line of sight needs to be comparable to the Doppler-broadened line width, which is the same in both clouds. If this is the case, radiation at some frequencies will  be amplified along the line of sight through a medium that combines material from both clouds. If the centers of the clouds pass close to each other along the line of sight, the likely result is a maser flare; see, for example, \citet{Lekht09}. The object we observe in \cep\ would, in this scenario, be either a pre- or postflare object, depending on whether the clouds are approaching, or separating from, their minimum line-of-sight separation. Multi-epoch observations would be necessary to test this model via proper motion analysis. At any frequency in the spectrum of the overlapping clouds, a ray amplifying through the overlapping region will pass through an optical depth $\tau_1$ of material from the first cloud and $\tau_2$ from the second cloud, with a resulting spectrum as shown, for example, in Fig.~9 of \citet{Lekht09}. We note that the differently shifted central response frequencies of the two clouds imply that the greatest optical depth, at a particular frequency, does not in general correspond to the greatest combined path length through the clouds, even if the lengths are otherwise identical. Comparison with our Fig.~\ref{fig:f6}, lower panel, suggests that our pair of clouds would be somewhat less overlapped than the Lekht \etal\ examples. Also, the model of only a pair of approximately spherical clouds does not naturally explain the variability pattern of the Pushchino monitoring. A more realistic model may involve nonspherical clouds or more overlapping clouds. In fact, this brings us close to the turbulence hypothesis discussed in the next section but without a pronounced turbulent vortex. 

\subsection{Structures in a Turbulent Flow} \label{sec:turbulence} 

In  the third dynamical model, we consider the case of turbulent vortices shed from the dense gas formation. Vortex formation, shedding, and evolution in the flow over dense obstacle are widely discussed in the literature, [e.g., \citet{Lienhard11, Blewins90, Loytsansky70}]. The turbulent motions have different regimes that are described by a set of dimensionless numbers (criteria). The corresponding regime of an unestablished flow is usually characterized by the Strouhal number, $St$. Expressed in observational parameters, it is equal to the ratio of characteristic scale, $R$, to the product of a characteristic speed, $v$, and characteristic time, $\tau$, i.e.,  $St=R/(\tau v)$. (Note: The Strouhal number is often defined as $St^{-1}$.) This number represents the ratio of the local velocity derivative to the convective derivative in the Navier--Stokes equation. Thus, this number describes the ability of the flow to form persistent turbulent vortices. The basic property of this criterion is that the Strouhal number $St$ has values from about 0.2 to about 0.3 for a wide range of Reynolds numbers, $Re$ [see mentioned textbooks, report by \citet{Roshko54}, and relatively recent experimental study by \citet{Shi11} and theoretical study by \citet{Ponta04}]. In order to facilitate discussion of our observations, we write $St$ as
\begin{equation}
St=57R_{\rm AU}/(\tau_m v_{\rm kms})~~,
\label{eq:eqstrouhal}
\end{equation}
where $\tau_m$ is the time in months, $v_{\rm kms}$ is the velocity in \kms, and $R_{\rm AU}$ is the spatial scale in astronomical units.

Pushchino monitoring shows that the 0.6\kms\ feature in \cep\ at the time of our observations experienced a rather strong flare, which is not likely to be periodic. Figure~\ref{fig:f4} shows that the flare lasted for about eight months and had two peaks at slightly different velocities. These peaks may correspond either to the full cycle of a single vortex rotation or to formation of two different vortices. 

Under the single-vortex hypothesis, the two maser spots correspond to two edges of the vortex. To estimate the Strouhal number, we adopt $\tau_m=16\pm6$, twice the lifetime of the 0.6\kms\ maser flare. We doubled the lifetime because (1) the full cycle of rotation implies that emission peak returns to the same velocity and (2) the arc in the position velocity dependence of the 0.6\kms\ feature (see Fig.~\ref{fig:f4}) suggests that it lasts for about half of a full cycle. Further, we assume that the component velocity difference 0.54\kms\ corresponds to a velocity difference of the edges of the vortex. Under this assumption, the characteristic velocity should be half of this value, i.e., $v_{\rm kms}=0.27$, and the measured value of the separation of the two components $R_{\rm AU}=0.11$ (160\muas) should be about the vortex diameter. The resulting Strouhal number is $St\sim1.5$, which is out of the normal range even for the cases of very high Reynolds numbers \citep{Green95, Schewe83}. Hence, we consider the hypothesis of the single turbulent vortex to be unlikely. 

In the two-vortex interpretation, each maser spot represents a vortex that forms in the wake of an obstacle in an outflow (von~K\'arm\'an street vortices). In our case, the line between the maser spots corresponds well to the axis of the outflow observed by \citet{Chibueze12}, and we consider this outflow as the progenitor of vortex formation. It is possible that the obstacle is associated with HW3diii. Subsequent vortices in the street rotate in the opposite sense. The density of the vortices decreases with the distance from the obstacle, so the dense gas responsible for the bright maser emission is present only in close proximity to the obstacle, so we observe only the first two. Vortex shedding has the following phases: (1) formation of one vortex with a component of velocity toward the flow axis on one side of the obstacle, (2) formation of another vortex with a component of its velocity toward the flow axis on the other side of the obstacle (at which stage we observe two dense vortices moving toward the flow axis from opposite sides), (3) the vortices approach the flow axis and start moving along the flow (in the meantime, a new vortex starts forming). When the obstacle is not symmetric, the vortices formed on one side of the obstacle can be denser, bigger, and, hence, brighter in maser emission. This model is consistent with the Pushchino monitoring results in Fig.~\ref{fig:f4} under the hypothesis that strong flares correspond to vortex formation, and we observe these structures moving along the flow axis. We should then observe two scales: the larger scale corresponds to vortex separation, or obstacle size (about 0.11~AU in our case), and the smaller scale, to two vortices with opposing rotation that manifest themselves at the highest angular resolution (our observed unresolved structures). Turbulence would therefore dissipate on scales much smaller than 0.11~AU in this region. Temporally, the period of the vortex shedding will correspond to half of the time difference between the strong flares, so about two months for the data in Fig.~\ref{fig:f4}; its characteristic velocity is about 10\kms, from typical proper motions measured by \citet{Chibueze12}, and the characteristic size is about 0.11~AU. These parameters give $St=0.3$, a plausible value for a turbulent flow in the interstellar medium. The hypothesis of a pair of turbulent vortices formed by an obstacle in the flow is therefore consistent with both the RadioAstron and Pushchino data from Fig.~\ref{fig:f4}.

\section{Conclusion} \label{sec:conclusion}

We have investigated the structure of a single maser ``spot" in the \cep\ region. We found that the maser spot had a total extent of about 400\muas. It is threaded by a magnetic field of less than 120~mG. The substructure is undoubtedly complex, but it includes two prominent structures separated by 160\muas, which contain about 13\% of the flux. The high contrast suggests that they may be unsaturated lines of sight. They may correspond to a pair of turbulent eddies shed by an obstacle in a flow, i.e., a~K\'arm\'an vortex street with Strouhal number of about 0.3, to objects bound in orbit by a planetary size mass, or to individual filaments or overlapping spherical clouds.

We note that the current study lacks information on the intermediate baselines, which are essential for accurate image recovery. Involvement of the High-Sensitivity Array (HSA) or full VLBA in observations of \cep\ in combination with RadioAstron would help to elucidate whether we have resolved the smallest scale of the turbulence, which is a basic parameter for understanding the evolution and structure of the interstellar medium of star-forming regions. Observations of flares should be conducted at intervals of a few months to determine their temporal and spatial characteristics.

\bigskip

 We thank  Vladimir Kostenko, Vyacheslav Avdeev and Pyotr Voitsik for help with correlation and calibration issues and Mark Reid for helpful suggestions on the manuscript. The RadioAstron project is led by the Astro Space Center of the Lebedev Physical Institute of the Russian Academy of Sciences and the Lavochkin Scientific and Production Association under a contract with the Russian Federal Space Agency, in collaboration with partner organizations in Russia and other countries. Partly based on observations performed with radio telescopes of IAA RAS (Institute of Applied Astronomy of the Russian Academy of Sciences). Partly based on observations with the Noto telescope operated by the Istituto di Radioastronomia di Bologna. Partly based  on observations with the 40-m radio telescope of the Yebes Observatory of the IGN (National Geographic Institute of Spain). Technical support was received from the National Space Facilities Control and Test Center. Results of optical position measurements of the Spektr-R spacecraft (the platform for the RadioAstron Space Radio Telescope) by the global MASTER Robotic Net, ISON collaboration, and the Kourovka Astronomical Observatory of the UrFU (Ural Federal University) were used for spacecraft orbit determination in addition to mission facilities.  This work was supported in part by the Ministry of Education and Science (the basic part of the State assignment, RK no. AAAA-A17-117030310283-7) and by the Act no. 211 of the Government of the Russian Federation, agreement 02.A03.21.0006. 

\facilities{RadioAstron Space Radio Telescope (Spectr-R), Yebes Radio Observatory (National Geographic Institute of Spain), Noto Radio Observatory (Bologna Institute of Radio Astronomy), and Zelenchukskaya Radio Observatory}

\software{ASC software correlator \citep{Likhachev17}, PIMA \citep{Petrov11}, AIPS \citep{vanMoorsel96}}

\vfill\eject

\end{document}